# Optimized designs for telecom-wavelength quantum light sources based on hybrid circular Bragg gratings

Lucas Rickert, Timm Kupko, Sven Rodt, Stephan Reitzenstein, and Tobias Heindel*

*Institut für Festkörperphysik, Technische Universität Berlin, Hardenbergstraße 36, D-10623 Berlin, Germany*
*Corresponding author: tobias.heindel@tu-berlin.de

**Abstract:** We present a design study of quantum light sources based on hybrid circular Bragg Gratings (CBGs) for emission wavelengths in the telecom O-band. The evaluated CBG designs show photon extraction efficiencies > 95% and Purcell factors close to 30. Using simulations based on the finite element method, and considering the influence of possible fabrication imperfections, we identify optimized high-performance CBG designs which are robust against structural aberrations. In particular, full 3D simulations reveal that the designs show robustness regarding deviations of the emitter position in the device well within reported positioning accuracies of deterministic fabrication technologies. Furthermore, we investigate the coupling of the evaluated hybrid CBG designs to single-mode optical fibers, which is particularly interesting for the development of practical quantum light sources. We obtain coupling efficiencies of up to 77% for off-the-shelf fibers, and again proof robustness against fabrication imperfections. Our results show prospects for the fabrication of close-to-ideal fiber-coupled quantum light sources for long distance quantum communication.

Quantum light sources emitting indistinguishable single-photon or entangled-photon states are key building blocks for future photonic quantum technologies [1-3], with applications ranging from quantum communication [4-6] to quantum computing [7-9]. In this context, quantum light sources based on epitaxial semiconductor quantum dots (QDs) are of particular interest, due to the availability of deterministic nanofabrication techniques [10-13] and the possibility to engineer and tailor photonic devices to specific needs [14]. Although the performance of QD-based quantum light sources has rapidly improved during the last decade, it remains challenging to meet all requirements set by applications in photonic quantum technologies in a single device concept. Micropillar cavities [15], for instance, are well suited for achieving large photon extraction efficiencies due to high β-factors and strong Purcell enhancement [16-18]. The underlying working principle, however, is intrinsically narrowband, hindering schemes which require the collection of photons from multiple, spectrally separated excitonic states [19,20]. Photonic microlenses deterministically fabricated above pre-selected QDs [21], on the other hand, provide enhanced photon extraction in a broad spectral range. The achievable extraction efficiencies exceed 50% at large numerical apertures, but do not reach the level of micropillar cavities and the microlens geometry does not provide reasonable Purcell enhancement. Devices based on circular Bragg gratings (CBGs) simultaneously offer high extraction efficiencies and Purcell enhancement [22]. Embedded in free-standing semiconductor membranes, however, they are challenging to fabricate.

Very recently, results on quantum light sources based on hybrid CBGs [23] with embedded QDs attracted much interest. Using these devices, benchmarks have been reported experimentally for the entanglement fidelity, the photon indistinguishability, the single-photon extraction efficiency and entangled-photon-pair collection efficiency [24,25]. These hybrid CBG devices seem so far unrivaled in all relevant optical and quantum optical properties, while benefitting from a potentially less fragile processing technology as they do not require free-standing membranes compared to non-hybrid CBGs. The emission wavelength, however, has up till now focused on short wavelengths of 780 nm and 880 nm, respectively. For applications in long-distance quantum communication, however, quantum light sources operating at telecom wavelengths are highly desirable, due to the low optical attenuation of silica fibers at these wavelengths. In addition, a detailed understanding how the CBG geometry affects the optical performance of fabricated devices is missing.

In this work, we perform an in-depth analysis on the design parameters of hybrid CBG-based quantum light sources emitting in the telecom O-band. Using simulations based on the finite element method (FEM), we evaluate designs for GaAs-based CBGs with an embedded quantum emitter integrated on a broadband reflective mirror. The evaluated device designs show Purcell factors as high as 30, while maintaining extraction efficiencies of > 95% of the embedded dipole's power. Furthermore, by investigating the impact of common fabrication imperfections on the optical properties of the CBGs, we prove the robustness of the evaluated designs against structural aberrations including QD positioning errors within reported deterministic fabrication uncertainties in the 5-50 nm range. Finally, we address the coupling efficiency of selected CBG device designs directly attached to single-mode optical fibers, representing an important setting for the development of plug-and-play photonic devices. Our results show prospects for the realization of close-to-ideal quantum light sources at telecom wavelengths.

To investigate the influence of various geometrical parameters on the performance of hybrid CBGs in the telecom O-band, we employ FEM simulations in the frequency domain using the software package JCMsuite [26]. The simulations utilize geometries discretized with a non-uniform-mesh triangulation and, unless stated otherwise, exploit the rotational symmetry of the CBG-structure by reducing the computational problem to a two-dimensional cross-section, thereby reducing the computational efforts [27]. The simulation domain is surrounded by perfectly matched layer (PML) boundaries. For the simulations, the room temperature refractive indices were used for all materials at a wavelength of $\lambda = 1320$ nm, already accounting for an expected blue shift of the resonant modes (of about 10 nm) due to refractive index changes at cryogenic temperatures (<10 K) required for non-classical light generation with highest quantum-optical quality [28]. The quantum emitter (QD) inside the structure is modeled as a TE dipole source at the center of the central disc of the CBG. The collection efficiencies of the evaluated designs are quantified by the dipole power collection efficiency (DCE), corresponding to the integrated power in the far-field emitted into NA=0.8 divided by the total power of the embedded dipole. For a better comparison with other reports [23], we also calculate the photon collection efficiency (PCE) defined as the dipole power emitted into NA = 0.8 divided by the power emitted into the upper half space.

Figure 1(a) shows a schematic cross-section of the hybrid CBG device, where the GaAs ($n_{GaAs} = 3.3885$) center disc and the grating are located on a silicon dioxide ($SiO_2$, $n_{SiO2} = 1.4500$) layer, combined with a gold (Au, $n_{Au} = 0.3970$, $k_{Au} = 8.9523$) layer for upward reflection of photons emitted into the lower half-space. The design parameters of the device are the thicknesses $t(Au)$, $t(SiO_2)$ and $t(GaAs)$ of the Au, the GaAs and the $SiO_2$ layer, respectively, the radius $R$ of the inner disk, the period length $P$ and gap-width $W$ of the surrounding Bragg-grating and the number of grating rings. Performing a detailed parameter study to optimize the extraction efficiency and the Purcell enhancement (considering the targeted operation wavelength), we found the following set of design parameters: $t(Au) = 100$ nm, $t(SiO_2) = 300$ nm, $t(GaAs) = 240$ nm, $R = 550$ nm, $P = 500$ nm, $W = 160$ nm and 11 rings forming the CBG. A top view of the corresponding hybrid CBG structure and the near-field profile of the supported fundamental optical mode are depicted in Figs. 1(b) and 1(c), respectively. The far-field intensity distribution of the same optical mode is shown in Fig. 1(d) for a NA of 0.8. Fig. 1(e) presents the Purcell factor $F_P$ and the collection efficiencies of the optimized CBG device. The DCE into NA = 0.8 reaches values exceeding 95% in this case. Noteworthy, although objectives with NA of 0.8 are commercially available for the O-band spectral range, even for NA = 0.4 this CBG device shows a DCE of 88% due to its high directionality (see Fig. 5 in the Supplementary Material). The PCE even reaches 98% for this NA, showing the effectiveness of the bottom gold mirror. The corresponding Purcell factor is close to 30 at the operation wavelength with a bandwidth (full width at half maximum) of $\Delta\lambda = 3.3$ nm.

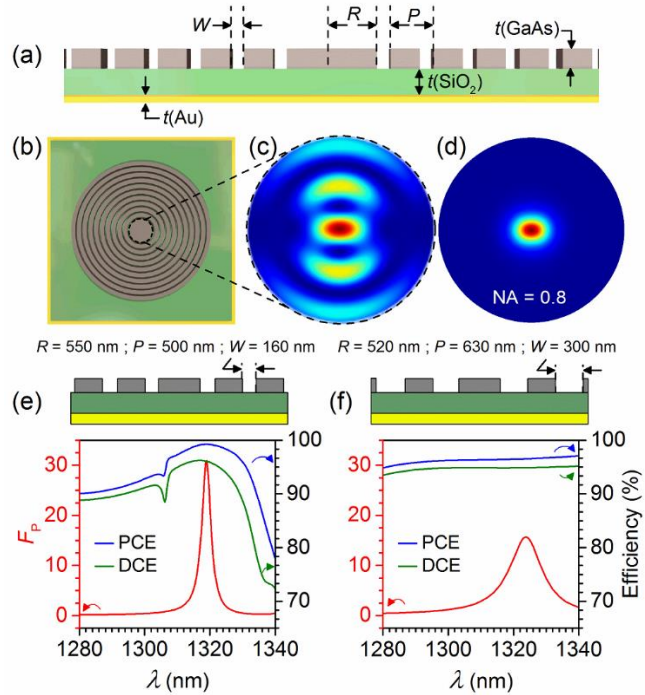

Fig. 1. (a) Schematic of the cross-section of a hybrid CBG device and relevant design parameters. (b) and (c) Top view of the CBG device and near-field intensity distribution of the supported optical mode inside the central CBG disc. (d) Far-field with NA = 0.8 showing the high directionality of the mode. (e) and (f) Simulated Purcell factor $F_P$, photon collection efficiency (PCE) and dipole power collection efficiency (DCE) as a function of wavelength $\lambda$ for a CBG design with narrowband and broadband characteristics, respectively.

Interestingly, our design study revealed that it is possible to achieve similar high performance in terms of the DCE for other sets of parameters, which might be easier to achieve with currently available in-situ lithography techniques. Figure 1(f) exemplarily presents simulations for a CBG device with a gap-width $W = 300$ nm, which is almost a factor of two enlarged compared to the device in Fig. 1(e). Optimization of the remaining parameters to $R = 520$ nm and $P = 630$ nm for maximum Purcell enhancement and extraction efficiency results in excellent performance in terms of DCE-values above 95% with operation at about $\lambda = 1323$ nm. Although the maximum Purcell factor of ~15 is somewhat lower in this case, the Purcell enhancement extends over a broader spectral range ($\Delta\lambda = 11.9$ nm). This is important for applications, such as the generation of polarization entangled photon-pairs via the biexciton- exciton (XX-X) radiative cascade in QDs. Here, the relevant excitonic states, i.e. X and XX, have a spectral separation of a few nanometers in case of InGaAs/GaAs QDs emitting in the telecom O-band [29]. In addition, the DCE shows an almost wavelength independent behavior, making the device more robust for applications such as the generation polarization-entangled photon pairs. Since the larger gap-width promises to be less demanding for lithography processes, we expect this large-$W$ design most suitable for device fabrication. The device's design parameters can be systematically changed in the following way: An increase in gap-width $W$ leads to a blue shift of the operation wavelength of maximum $F_P$. An increase in ring-size (by increasing $P$, while keeping $W$ constant) causes a red-shift in operation wavelength of maximum $F_P$. This is due to the effective refractive index experienced by the mode, and thus by changing $P$ and $W$ with respect to each other, the operation wavelength can be systematically tuned to a desired value. Similarly, the effects of $P$ and $W$ on the out-coupling properties are systematic, where an increase in $P$ (or decrease in $W$) leads to a blue-shift of the operation wavelength compared to the spectral regions with high DCE-values. A change of the parameters in the opposite fashion causes an opposite shift of operation wavelength and DCE-values. Detailed information on the influence on the optical properties of each parameter, including the number of grating rings, are documented in the Supplementary Material, Fig. 5-7.

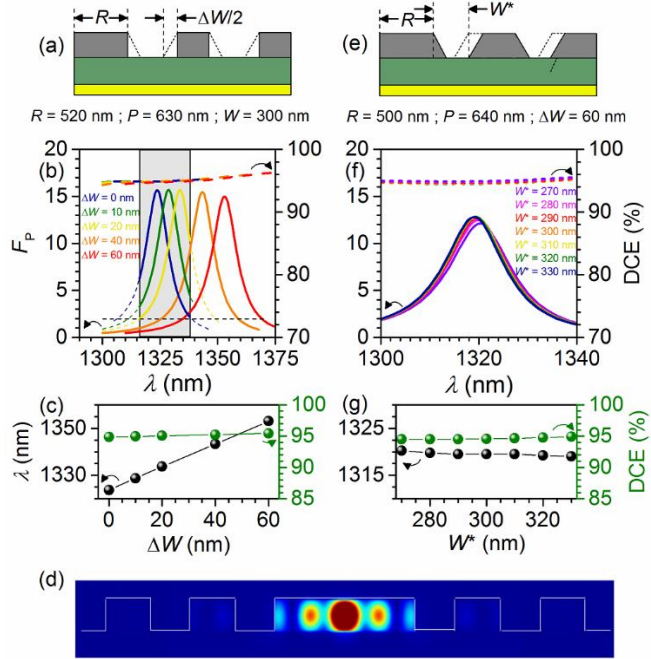

Fig. 2. (a) Schematic cross-section of a hybrid CBG device considering fabrication imperfection resulting in narrower gaps with tilted side walls and reduced width $W-\Delta W/2$. (b) $F_P$ and DCE as a function of wavelength for different $\Delta W$. The gray shaded spectral region indicates $F_P > 2$ (dashed horizontal line), if the side-wall imperfection is limited to $dW = 6.67\%$ ($\Delta W = 20$ nm). (c) Operation wavelength $\lambda$ (at maximum $F_P$) and DCE as a function of $\Delta W$. (d) Cross-sectional view of the near-field intensity distribution inside the CBG. (e) Schematic of the same fabrication imperfection as in (a) but with fixed central disc and variable gap-width $W^*$. (f) $F_P$ and DCE as a function of wavelength for various $W^*$. (g) $\lambda$ and DCE at maximum $F_P$ as a function of $W^*$.

Next, we consider fabrication imperfections potentially affecting the optical performance of hybrid CBG devices. A typical issue in the device fabrication is an insufficient etching of the grating gaps, resulting in tilted side walls of the grating. This tilt effectively leads to a reduced gap-width W and an enlarged diameter of the central disc. Such problems arise for instance from insufficient development (under-development) during lithography, which negatively affects subsequent etch steps. Figure 2(a) illustrates such imperfections with a target gap-width of $W = 300$ nm and a variable tilt of the sidewalls described by $\Delta W$. In this notation, a discrepancy of $\Delta W = 60$ nm corresponds to an imperfection of $dW = \Delta W/W$ of 20% or a deviation of about 8° from a vertical side wall. As seen from Figs. 2(b) and 2(c), the operation wavelength is significantly affected by the fabrication imperfections, while the DCE remains at a high level (about 95% for NA = 0.8) almost independent of the tilt. The gray shaded region in Fig. 2(b) corresponds to a spectral window of 23 nm over which a moderate Purcell enhancement with $F_P = 2$ can be achieved, despite the shift in operation wavelength, if the imperfections can be limited to $dW = 6.67\%$. Moreover, a slight decrease in the maximal $F_P$ and an increase in the bandwidth of the Purcell enhancement is observed with increasing tilt. The high sensitivity of the operation wavelength to the tilt is attributed to the fact that the latter effectively leads to an enlarged size of the central disc (increased R). The size of the central disc in turn governs the confinement of the optical mode, due to relatively high near-field intensities close to the disc edge, as seen in Fig. 2(d). A change in disc size causes a noticeable change in the effective refractive index experienced by the optical mode, resulting in a shift of the operation wavelength. To verify this explanation, we performed simulations with fixed disc radius and tilt ($R = 500$ nm, $dW = 20\%$), but variable effective gap-width $W^*$. Figures 2(e)-2(g) show the influence of $W^*$ on $F_P$, DCE and the operation wavelength $\lambda$. Varying $W^*$ from -30 nm to +30 nm relative to the targeted value $W = 300$ nm, this design shows a robust behavior with almost independent values for $F_P$ of about 15, DCE of about 95% and operation wavelengths close to 1320 nm. Consequently, if the size of the central disc can be controlled well enough (within $\Delta R = \pm 20$ nm), the constraints for the remaining design parameters are relaxed, and Purcell enhancement is achieved at the desired wavelength. Even if the disc size is controlled less precise, the out-coupling efficiency remains largely unaffected by these imperfections. That means that a high optical performance can be achieved, despite of structural aberrations in terms of sidewall tilts and varying gap-widths $W$. This finding is very promising for the robust fabrication of near-optimal telecom wavelength quantum light sources based on hybrid CBG devices.

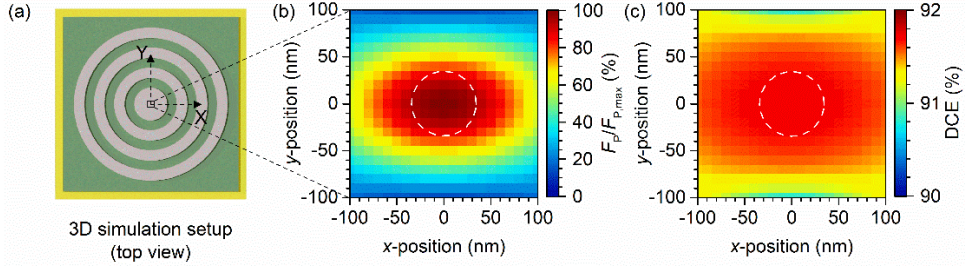

Figure 3. (a) Schematic of the 3D simulation setting to investigate the hybrid CBG devices with a dipole emitter deviating from the central position. (b) and (c) $F_P$- and DCE-values for different $x$- and $y$- emitter positions in the disc. The white circles indicate a deviation of 34 nm from the center.

Another possible fabrication imperfection is the lateral displacement of the quantum emitter with respect to the center of the device. To investigate the influence of this imperfection, full 3D simulations of CBG devices with off-center dipole emitters were performed (cf. Fig. 3(a)). Taking advantage of the fact, that the broadband design from Fig. 1(f) requires only 3 Bragg rings as compared to 11 rings for the narrowband design of Fig. 1(e), allowed us to significantly reduce the size of the computational domain to 6.04 µm × 6.04 µm × 0.84 µm, resulting in convenient simulation times of about 20 minutes (per emitter displacement) despite the significantly increased computational costs of the 3D setting. The broadband design with only 3 rings still shows $F_P$-values close to 15 and DCE-values close to 92% (cf. Fig. 5(a) in Supplementary Material), Fig. 3(b) and 3(c) show the Purcell factor (normalized to its maximum $F_{P,max} \sim 15$) and DCE-values for lateral emitter positions on a 200 nm × 200 nm grid around the center of the device. A $x$- and $y$-displacement of the dipole emitter of 34 nm, which corresponds to the reported positioning accuracy of in-situ electron-beam lithography [30], results in a reduction of $F_P$ by only 5.4% in $x$-direction and 15.6% in $y$-direction, respectively, where the asymmetry results from the linearly polarized dipole chosen to mimic a quantum emitter. The out-coupling efficiency is even more robust, with DCE-values for NA of 0.8 reduced by only 0.02% and 0.05% for a 34 nm displacement in $x$- and $y$-direction, respectively. Considering the simulation results above and the fact, that device positioning accuracies down equal or better than 50 nm [30-32] have been already demonstrated experimentally, the broadband design shows robust behavior also for a finite lateral displacement of the quantum emitter inside the device. We also want to emphasize, that the observed asymmetry in the $x$- and $y$-dependency originates from the polarized emission of the embedded dipole source. Quantum emitters such as QDs typically emit a mixture of both polarizations and the mode near-field in this case would be circularly symmetric and extents its field further into the $y$-direction. Noteworthy, the observed behavior of $F_P$ follows the distribution of the electric field-intensity of the confined optical mode. Since the spatial extend of the optical mode is larger for longer wavelengths, we therefore expect in general a higher robustness for a lateral emitter displacements of CBG devices operating at telecom-wavelengths compared to devices optimized for shorter wavelengths.

For the practical use of quantum light sources in photonic quantum technologies, a fiber-coupling of the emission is highly beneficial. Therefore we investigate the direct coupling of telecom-wavelength hybrid CBG devices to optical single-mode fibers using FEM simulations in a rotational symmetric setting as introduced earlier. We considered two different types of commercially available fibers: first 980HP with a core diameter of $d_{core} = 3.6$ µm and refractive indices for core and cladding of $n_{core} = 1.460$ and $n_{cladding} = 1.447$ (at $\lambda = 1320$ nm), and second SMF28, assuming $d_{core} = 4.07$ µm, $n_{core} = 1.452$ and $n_{cladding} = 1.447$. Calculations were carried out by assuming vacuum or UV adhesive (NOA81, $n = 1.56$ [33]) inside the gap between fiber and CBG perfectly aligned in axial direction. Using adhesives, a permanent and stable coupling can be achieved between a photonic device and a fiber [34,35]. The schematic in Fig. 4(a) shows the setting including the optical fiber and the hybrid CBG device, which was used to simulate the mode coupling efficiency (MCE) defined as the fraction of the dipole power which is coupled to the two-fold degenerate fundamental mode of the fiber [27]. Figure. 4(b) presents the MCE as a function of the distance d between fiber and CBG for the narrowband and broadband design presented in Figs. 1(e) and 1(f), respectively. For the coupling to 980HP fiber in vacuum, a maximum MCE of 76.0% and 50.3% is observed for the narrowband and broadband design, respectively. With UV adhesive, a maximum MCE of 54.6% is found for the narrowband design. In this case the radius of the central disc was reduced from $R = 550$ nm to $R = 540$ nm to account for the red shift induced by the present adhesive due to its higher refractive index compared to air, and to achieve λ close to 1320 nm. All three simulated curves show a periodic oscillation of MCE as a function of d, due to interference of the emitted light field inside the gap region. Moreover, both designs show a plateau-like behavior of the mean MCE at small distances $d < 4$ μm, followed by an almost linear decrease. This behavior is similarly observed for the simulation with adhesive, the MCE, however, is less sensitive to $d$ in this case. The plateau reflects the high directionality of the emission of the hybrid CBG device, and is promising for the experimental realization of direct fiber-coupling. To optimize the fiber-coupling efficiency for both designs, Fig. 4(c) presents the maximum MCEs (at $d = 1350$ nm) under variation of W around its initial values. For the narrowband design in vacuum, a slightly higher MCE of 76.9% can be achieved (at $W = 170$ nm) for the 980HP fiber, while a significant improvement is possible for the broadband design which shows up to 69.6% for this fiber-type in vacuum (at $W = 345$ nm). With UV adhesive, the narrowband design shows a maximum MCE of 61.4% at $W = 180$ nm ($d = 650$ nm).

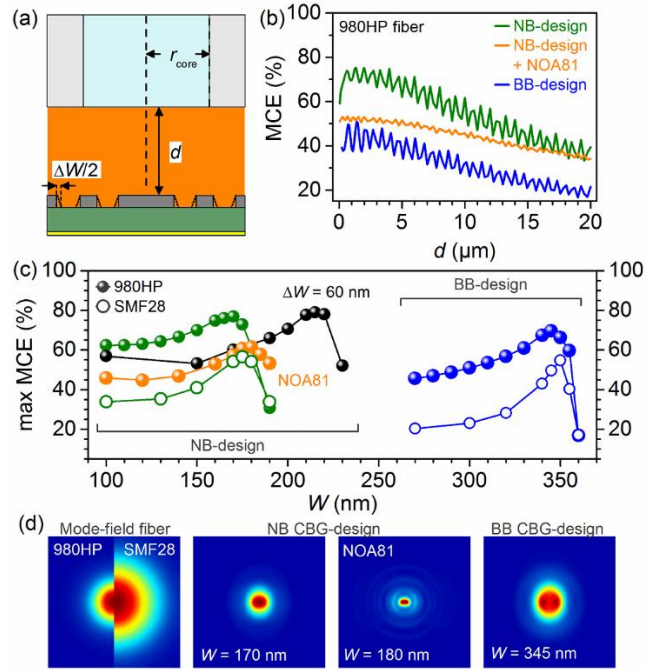

Figure 4. (a) Schematic of the simulation setting to investigate the fiber-coupling of hybrid CBG devices emitting in the telecom O-band. (b) Mode coupling efficiency (MCE) to a 980HP fiber as a function of distance $d$ between fiber and CBG for the narrowband (NB-) and broadband (BB-) designs from Figs. 1(e) and 1(f) for vacuum and adhesive (NOA81) between fiber and CBG device, respectively. (c) Maximum achievable MCE under variation of the gap-width $W$ around its initial value for different narrowband (NB-) and broadband (BB-) designs, considering the coupling to 980HP and SMF28 fiber. For the narrowband design, the MCE is also calculated with an adhesive (NOA81) in-between fiber and CBG. For the NB-design with adhesive, the central disc radius was reduced to $R = 540$ nm to account for a redshift caused by the adhesive with higher refractive index compared to air. (d) 2D fiber-mode profiles and CBG-mode profile at the distance of maximum MCE for narrowband and broadband designs.

In addition, Fig. 4(c) depicts results obtained from a design with imperfect sidewalls ($\Delta W = 60$ nm) coupled to the 980HP fiber in vacuum. Even in this case, coupling efficiencies close to 80% are reached, confirming that also the fiber-coupling of hybrid CBG devices is robust against fabrication imperfections. Additionally, calculations for the coupling of CBGs to SMF28 fibers in vacuum are presented for the narrowband and broadband design under variation of W (assuming ideal sidewalls). In this case, the MCE reaches values of 56.4% (at $W = 175$ nm) and 54.8% (at $W = 350$ nm). In general, all curves for the MCE show a less sensitive behavior for shorter gap-widths $W$ (for about 60 nm), but an abrupt degradation towards longer $W$. We attribute the observed dependencies of the MCE for different $W$ to the degree of overlap between the near-field of the CBG at the respective distance and the Gaussian fiber modes. As suggested by the field intensity distribution for $W = 170$ nm of the narrowband design in Fig. 4(d), highest MCE is achieved for an almost symmetrical Gaussian shape of the near-field distribution in X- and Y- direction. For $W = 345$ nm of the broadband design in Fig. 4(d), the field distribution shows larger asymmetries and deviations from a Gaussian shape, resulting in a reduced mode-overlap between CBG and fiber. This mode-mismatch becomes even more pronounced for $W = 190$ nm and $W = 360$ nm, since volume and lateral extend of the CBG mode become large compared to the fiber mode (see Fig. 8 in the Supplementary Material). Importantly, the coupling to the 980HP fiber is significantly more efficient (by a factor of 1.2 to 2.2 for specific $W$ for narrowband and broadband design) compared to the SMF28 fiber, which results from a better mode-matching for the smaller core-size fiber. This becomes even more important in the presence of UV adhesive, as the already highly directional CBG mode-field is further squeezed in lateral dimensions.

In summary, the predicted coupling efficiencies to single-mode fibers surpass values reported for other systems, including microlenses [27], photonic trumpets [36], and photonic crystal cavities [37], while being only marginally smaller compared to coupling efficiencies reported for tapered photonic crystal waveguides [38] and micropillar structures [35]. Further improvements in the coupling performance are expected, e.g. by optimizing the combined design parameters of hybrid CBG devices using optimization algorithms [39] and specialty fibers with tailored geometry and index profiles. Last but not least, the influence of surface roughness (random disorder) and lateral asymmetries on the performance of CBG devices with and without fiber-coupling is interesting for future studies.

In conclusion, we presented a detailed design study for the fabrication of quantum light sources based on hybrid CBG devices operating in the telecom O-band. Using FEM simulations, we evaluated optimized device designs for highest performance in terms of photon-extraction efficiency, enabling Purcell-Factors as high as 30 and dipole power extraction efficiencies exceeding 95%. Furthermore we showed, that the excellent performance of the proposed hybrid CBG designs can be maintained and optimized for technically less demanding feature sizes. Additionally, our designs proved to be robust against common fabrication issues in terms of imperfect sidewall etching and lateral displacement of the emitter's position inside the device. Finally, we demonstrated the prospects for a highly efficient and robust direct fiber-coupling of hybrid CBG devices, revealing mode-coupling efficiencies close to 80% for off-the-shelf single-mode optical fibers. Further improvements are anticipated using customized/tailored specialty fibers in combination with optimized CBG design parameters. Our results are encouraging towards the fabrication of near-optimal and practical quantum light sources at telecom wavelength for various applications in photonic quantum technologies.

**Acknowledgment:**
We acknowledge financial support of the German Federal Ministry of Education and Research (BMBF) via the project 'QuSecure' (Grant No. 13N14876) within the funding program Photonic Research Germany. We further acknowledge Sven Burger and Philipp-Immanuel Schneider from the company JCMwave for helpful discussions and support.

**Supplementary Material**

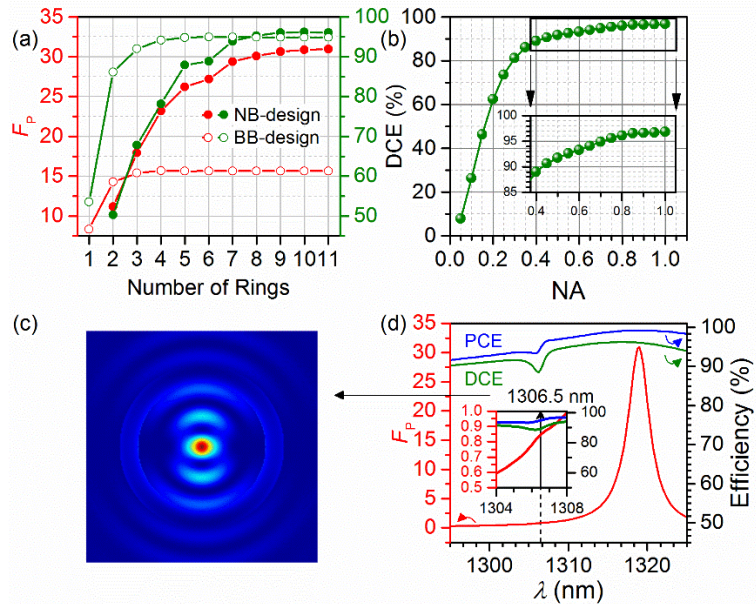

Figure 5 – complementing Fig. 1(e) and 1(f): (a) Maximum Purcell factor $F_P$ and dipole power collection efficiency DCE for the narrowband (NB)- and broadband (BB)- CBG device with varying number of rings. A $F_P$ close to 30 (above 15) and a DCE of over 90% is reached for devices with more than 6 rings (2 rings) for the NB-device (BB-device). (b) DCE of the NB-device at $\lambda = 1319$ nm as a function of the numerical aperture (NA). Due to the directional emission of the hybrid CBG devices, close to 90% of the dipole emission power is collected already for a NA of 0.4. (c) Near-field intensity distribution inside the circular Bragg grating (CBG) at a wavelength of $\lambda = 1306.5$ nm, at which a dip is observed in the dipole collection efficiency of the NB-device in (d). A closer look also reveals a local maximum in $F_P$ at the same spectral position, confirming that this substructure originates from another mode of the hybrid CBG device.

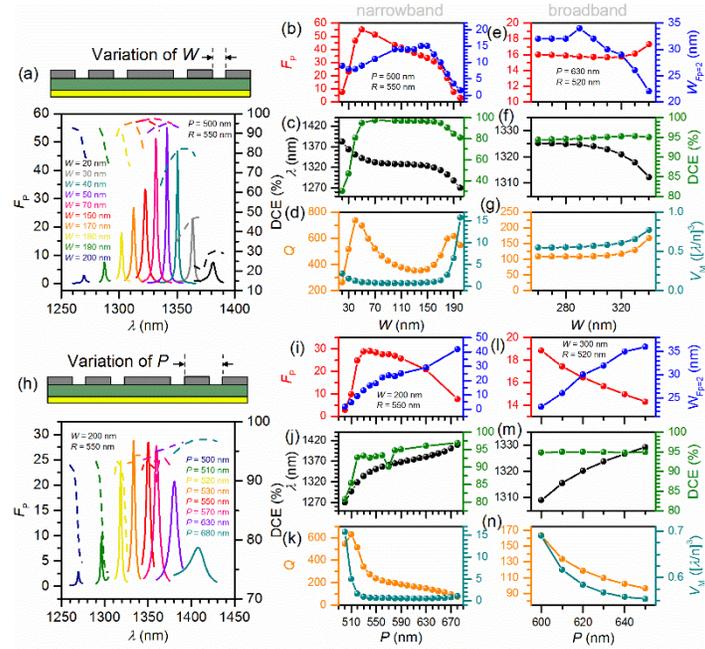

Figure 6 – complementing Fig. 1. Influence of gap-width $W$ of the CBG on the optical properties for the narrowband ($W$ = 160 nm) [(a)-(d)] and broadband ($W$ = 300 nm) [(e)-(g)] design, whose optical properties are displayed in Fig. 1(e) and 1(f), respectively. Influence of grating period $P$ on optical properties for the narrowband design [(h)-(k)] and broadband design [(l)-(n)]. $W$ = 100 nm (narrowband) and $W$ = 260 nm (broadband) result in near-field intensity distributions with very low intensity contributions in the grating regions compared to $W$ = 200 nm and $W$ = 340 nm, which show a much higher sensitivity of $\lambda$ on $W$ and $P$. As seen from (a) and (h), the Purcell enhancement follows a bell-shaped distribution for variations in $W$ and $P$, which can be correlated with the Q-factor $Q$ and mode volume $V_M$ exhibited by the mode for the specific design. For certain $W$, Purcell factors of 55 are obtained. Since the maximum Q-factor found is 800, we believe that the weak-coupling regime is valid for the entire investigated parameter range, although a very small mode volume of ~0.6 $(\lambda/n)^3$ ($\approx$ 0.034 µm³) is given. The higher $Q$, the spectrally narrower is the Purcell enhancement ($W_{F_P>2}$ corresponds to the spectral width with $F_P > 2$). As can be further seen from (a) and (h), decreasing $W$ or increasing the ring size (i.e. increasing $P$ while keeping $W$ constant) leads to a red-shift of the operation wavelength and maximum $F_P$. This simultaneously shifts the maximum $F_P$ to shorter wavelengths compared to the spectral regions with high DCE-values, and the operation wavelength can be aligned to an optimal out-coupling. An increase in $W$ or decrease in ring size has the opposite effect.

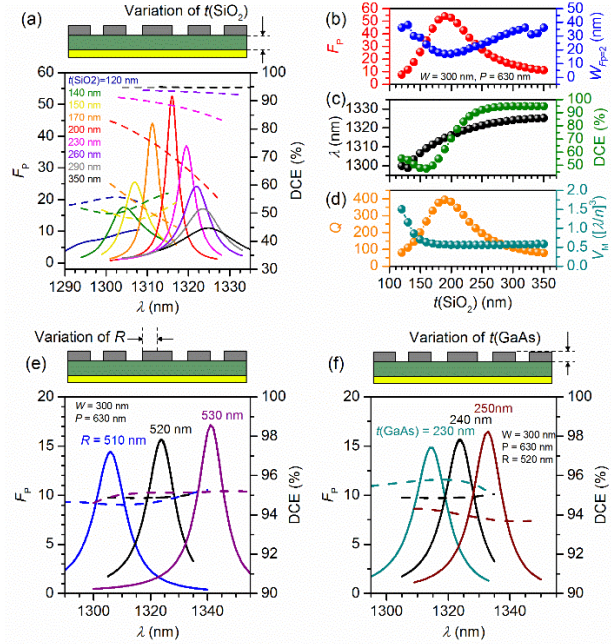

Figure 7 (a)-(d) Influence of the thickness $t(SiO_2)$ of the $SiO_2$ layer on the optical properties for the broadband design with $W = 300$ nm (cf. Fig. 1(f)). The Purcell enhancement follows again a bell-shaped curve with varied $SiO_2$ thickness. (b) and (d) show clearly that this change in $F_P$ arises from the influence of $t(SiO_2)$ on $Q$, whereas $V_M$ is constant over the varied thickness range, except for very thin dielectric layers. A maximum $F_P = 54$ can be obtained at $t(SiO_2) = 190$ nm. Regarding the origin of this resonance, we found that the 190 nm of $SiO_2$ form a $\lambda/2$ distance between the position of the dipole and the gold mirror, corresponding to $(1/2\, t(GaAs) \cdot n_{GaAs})/\lambda + (t(SiO_2) \cdot n_{SiO2})/\lambda$, $t(GaAs) = 240$ nm ; $n_{GaAs} = 3.3885$ ; $n_{SiO2} = 1.45$ ; $\lambda = 1314.5$ nm. This means, that the emitted light travels a $\lambda$-distance back and forth to the mirror and is able to constructively interfere with the emitter again. On the other hand, a $SiO_2$ thickness of at least 250 nm is required to achieve DCEs larger 90%. (e) and (f) Influence of the central disc radius $R$ and CBG membrane thickness $t(GaAs)$ on the optical properties. Increasing $R$ and $t(GaAs)$ both red-shift the operation wavelength, which can be used to fine-tune $\lambda$. The larger shift of $\lambda$ with $R$ is attributed to the high mode intensity at the edges of the disc.

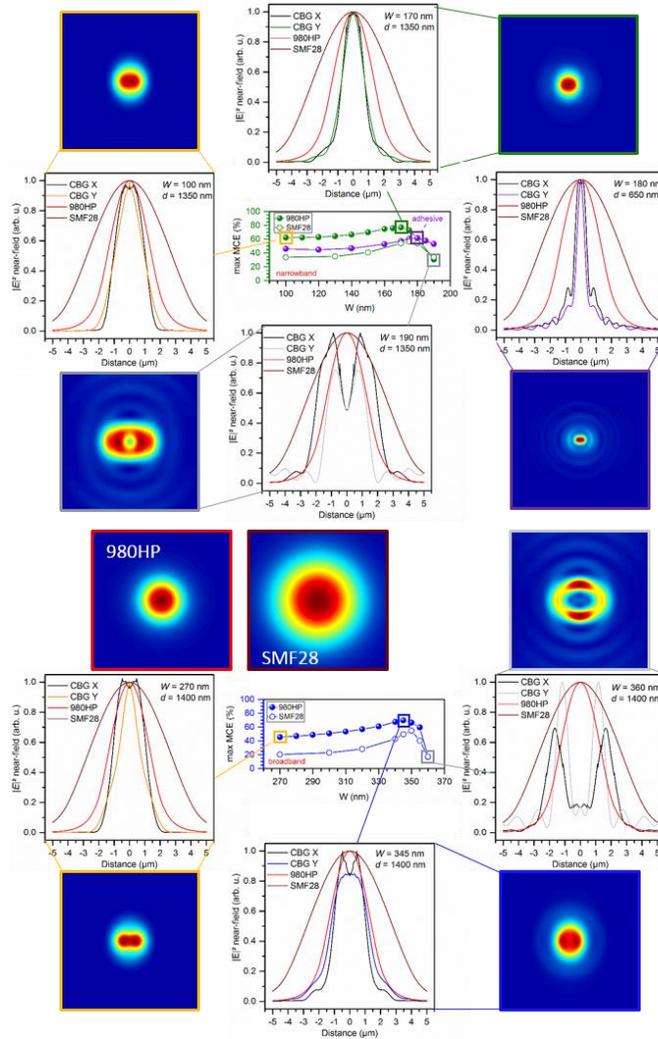

Fig. 8 – complementing Fig. 3: 980HP fiber-mode profiles and CBG field-intensity distributions at a distance *d* of maximum fiber-mode coupling efficiency for the narrowband and broadband designs with *W* around 160 nm and 300 nm. The intensity cross-cuts and 2D intensity distributions indicate, that highest coupling efficiency is achieved at optimum overlap of CBG field- and fiber-mode profile.